\documentclass[aps,reprint,pra,superscriptaddress,floatfix,
]{revtex4-1}  
\usepackage{graphicx,epsfig,epstopdf}
\usepackage[dvipsnames]{xcolor}
\usepackage{amsmath}
\usepackage{amsthm}
\usepackage{subfigure}
\usepackage{mathrsfs}
\usepackage{bm}
\usepackage{graphicx}
\usepackage{times}

\usepackage{upgreek}

\newtheoremstyle{query}%
{}{}
{\color{blue}}
{}
{\sffamily\bfseries}{:}{12pt}
{}
\theoremstyle{query}
\newtheorem{aq}{Author Query/Comment}

\newcommand{\baq}{\begin{aq}}
\newcommand{\eaq}{\end{aq}}

\begin{document}

\title{Experimental Realization of Nonadiabatic Holonomic Single-Qubit Quantum Gates\\ with Optimal Control in a Trapped Ion}

\author{Ming-Zhong Ai}
\affiliation{CAS Key Laboratory of Quantum Information, University of Science and Technology of China, Hefei 230026, China}
\affiliation{CAS Center For Excellence in Quantum Information and Quantum Physics,
	University of Science and Technology of China, Hefei 230026, China}

\author{Sai Li}
\affiliation{Guangdong Provincial Key Laboratory of Quantum Engineering and Quantum Materials, 
and School of Physics\\ and Telecommunication Engineering, South China Normal University, Guangzhou 510006, China}

\author{Zhibo Hou}
\affiliation{CAS Key Laboratory of Quantum Information, University of Science and Technology of China, Hefei 230026, China}
\affiliation{CAS Center For Excellence in Quantum Information and Quantum Physics,
	University of Science and Technology of China, Hefei 230026, China}

\author{Ran He}
\affiliation{CAS Key Laboratory of Quantum Information, University of Science and Technology of China, Hefei 230026, China}
\affiliation{CAS Center For Excellence in Quantum Information and Quantum Physics,
	University of Science and Technology of China, Hefei 230026, China}

\author{Zhong-Hua Qian}
\affiliation{CAS Key Laboratory of Quantum Information, University of Science and Technology of China, Hefei 230026, China}
\affiliation{CAS Center For Excellence in Quantum Information and Quantum Physics,
	University of Science and Technology of China, Hefei 230026, China}

\author{Zheng-Yuan Xue}\email{zyxue83@163.com}
\affiliation{Guangdong Provincial Key Laboratory of Quantum Engineering and Quantum Materials, 
and School of Physics\\ and Telecommunication Engineering, South China Normal University, Guangzhou 510006, China}
\affiliation{Frontier Research Institute for Physics, South China Normal University, Guangzhou 510006, China}

\author{Jin-Ming Cui}\email{jmcui@ustc.edu.cn}
\affiliation{CAS Key Laboratory of Quantum Information, University of Science and Technology of China, Hefei 230026, China}
\affiliation{CAS Center For Excellence in Quantum Information and Quantum Physics,
	University of Science and Technology of China, Hefei 230026, China}

\author{Yun-Feng Huang}\email{hyf@ustc.edu.cn}
\affiliation{CAS Key Laboratory of Quantum Information, University of Science and Technology of China, Hefei 230026, China}
\affiliation{CAS Center For Excellence in Quantum Information and Quantum Physics,
	University of Science and Technology of China, Hefei 230026, China}

\author{Chuan-Feng Li}\email{cfli@ustc.edu.cn}
\affiliation{CAS Key Laboratory of Quantum Information, University of Science and Technology of China, Hefei 230026, China}
\affiliation{CAS Center For Excellence in Quantum Information and Quantum Physics,
	University of Science and Technology of China, Hefei 230026, China}

\author{Guang-Can Guo}
\affiliation{CAS Key Laboratory of Quantum Information, University of Science and Technology of China, Hefei 230026, China}
\affiliation{CAS Center For Excellence in Quantum Information and Quantum Physics,
	University of Science and Technology of China, Hefei 230026, China}

\date{\today}

\begin{abstract}
Quantum computation with quantum gates induced by geometric phases is regarded as a promising strategy in fault-tolerant quantum computation, owing to its robustness against operational noise. However, because of the parametric restrictions in previous schemes, the main robust advantage of holonomic quantum gates is reduced. Here, we experimentally demonstrate a solution scheme, obtaining  nonadiabatic holonomic single-qubit quantum gates with optimal control in a trapped $^{171}\mathrm{Yb}^{+}$ ion  based on a three-level system with resonant driving, which also has the advantages of rapid evolution and convenient implementation. Compared with  previous geometric gates and conventional dynamical gates, the superiority of our scheme is that it is more robust against control amplitude errors, which is confirmed by the  gate infidelity  as measured by both quantum process tomography and random benchmarking methods. In addition, we  outline how nontrivial two-qubit holonomic gates can also be realized using currently available experimental technology. Thus, our experiment confirms the feasibility of this  robust and fast holonomic quantum computation strategy.
\end{abstract}

\maketitle


\section{INTRODUCTION}

Recently, the construction of   practical quantum computers has attracted much attention. However, any large-scale quantum system is inevitably subjected to the influence of control fields and the surrounding environment,  leading to  corruption of the desired quantum information. Thus, there is a need for robust and fast quantum information processing. Interestingly, both Abelian \cite{berry84} and non-Abelian \cite{zee84} geometric phases, which  depend only on the global properties of the evolution trajectories, have intrinsic robust features against certain local noises \cite{ps04,zhu05,ps12,mj12}. Moreover, given the decoherence of  quantum systems, nonadiabatic evolution \cite{AA87} is  preferred to  adiabatic evolution \cite{berry84,zee84}, which requires a  long run-time. In this context, high-fidelity quantum gates obtained in a nonadiabatic geometric way are  of great interest \cite{xbwang01,zhu02}, and therefore considerable attention has been devoted to  nonadiabatic geometric quantum computation \cite{gqc08}.

The non-Abelian geometric phase has the intrinsic property of noncommutativity, and thus it can naturally be used to realize universal quantum gates for  so-called holonomic quantum computation, which was originally proposed \cite{zanardi99,Duan01} and experimentally demonstrated \cite{toyoda2013realization,leroux2014} on the basis of adiabatic evolution. However, environmentally induced noise will impose a severe limitation on  gate performance, because of the limited  coherence time of a quantum system. To overcome this limitation,  nonadiabatic holonomic quantum computation (NHQC) schemes \cite{Sjoqvist2012,Xu2012} have been proposed, in which  the restriction to an adiabatic condition is removed. The pioneering NHQC schemes have been  extended  theoretically \cite{Xu2015,es2016,Herterich2016,Xue2017,xu20172, xu2017,Hong18,xu2018,nr2019} and   verified  experimentally  using superconducting circuits \cite{Abdumalikov2013,Xu18,Danilin18}, nuclear magnetic resonance (NMR) \cite{Feng2013,li2017,zhu2019},  nitrogen-vacancy centers in diamond \cite{Zu2014,Arroyo-Camejo2014,nv2017,nv20172,ni2018,kn2018}, and other approaches. However, owing to restrictions on the parameters  in the governing Hamiltonian, these NHQC schemes are subject to systematic errors \cite{Zheng16,Jing17}, which  removes  the main advantage of  geometric quantum gates.

Recently, theoretical schemes \cite{Liu19,  Li20} have been proposed to relieve the parametric constraints of previous NHQC schemes, and an arbitrary holonomic quantum gate  can thereby be achieved  in a single-loop evolution.   Following Ref.~\cite{Liu19},  a shortcut to non-Abelian geometric gates has been  demonstrated  experimentally \cite{yan2019} in a superconducting circuit  with off-resonance drives. However, for compatibility with pulse shaping, the detuning between the drive and the corresponding qubit should be time-dependent, and  furthermore, the detuning and the driving amplitudes should be precisely controlled in a correlated way, which is   difficult to achieve experimentally.  Therefore, the implementation of a robust NHQC (RNHQC) with resonant drives is a preferable experimental approach, since it only requires control of the two resonant driving fields.

In this paper, we present an experimental demonstration of nonadiabatic holonomic quantum gates in the RNHQC scheme \cite{Li20}  using a trapped $^{171}\mathrm{Yb}^{+}$ ion with a three-level configuration. To introduce optimal control for the holonomic gates, the evolution of the nonadiabatic holonomy is induced by modulating both the time-dependent amplitude and phase of a two-tone resonant microwave drive. In our realization, characterized by a random benchmarking  (RB) method, the obtained average gate fidelity is above 99\% and is  restricted mainly by the limited coherence times. Moreover, we demonstrate that our  holonomic gates are more robust against control amplitude errors than  both previous NHQC schemes and  conventional dynamical gates, under the same maximum drive amplitude. In addition, in combination with a nontrivial  nonadiabatic holonomic two-qubit gate,  universal RNHQC can be achieved in this trapped-ion setup using current technology. Thus, our experimental results demonstrate the possibility of RNHQC and are relevant for the development of practical large-scale quantum systems.

\begin{figure}
\includegraphics[width=8.5cm]{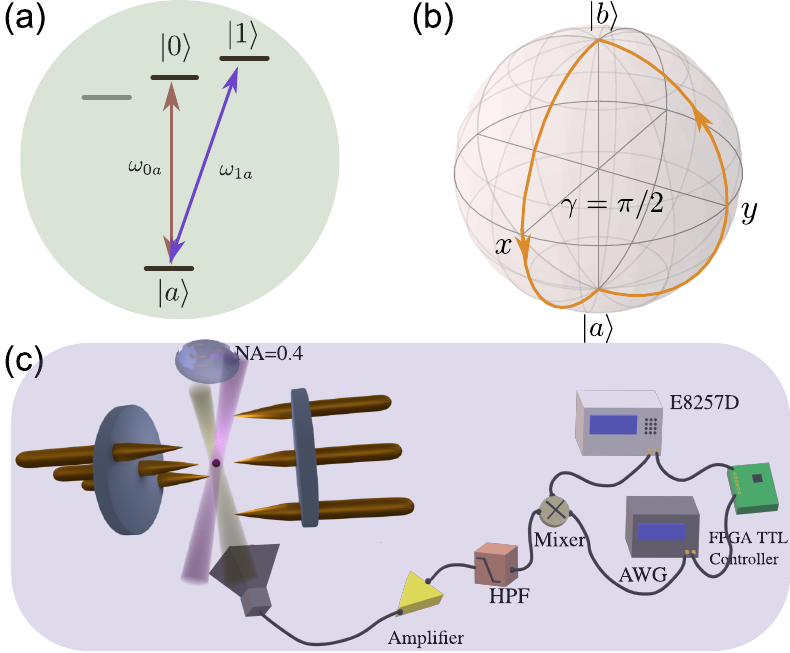}
\caption{Realization of arbitrary single-qubit nonadiabatic holonomic  gates. (a) Hyperfine energy levels of a $^{171}\mathrm{Yb}^{+}$ ion. The qubit states $|0\rangle$ and $|1\rangle$ are encoded in $|^{2}S_{1/2}, F=1,m_{F}=0 \rangle$ and $|^{2}S_{1/2}, F=1,m_{F}=1 \rangle$, respectively. Two microwave fields $\omega_{0a}$ and $\omega_{1a}$ are resonantly coupled with transitions  $|0\rangle \leftrightarrow |a\rangle$ and $|1\rangle \leftrightarrow |a\rangle$ to generate holonomic gates in the qubit subspace. (b) Illustration of the implemented  holonomic gates  in the Bloch sphere of  the $\{ |b\rangle, |a\rangle \}$ subspace. The  whole evolution is divided into two steps: first evolving from the bright state $|b\rangle$ to the auxiliary state $|a\rangle$ and then back with an additional phase. (c) Our simplified experimental setup. An ion is trapped in a needle trap that has a pair of rf electrodes and two pairs of dc electrodes. The microwaves generated from an AWG mix with microwave from a signal generator (E8257D), and then this mixed signal interacts with the ion through a microwave horn.}
\label{fig:device}
\end{figure}

\section{universal single-qubit gates}
We first address the realization of arbitrary holonomic single-qubit gates in the $\{|0\rangle,|1\rangle\}$ subspace on $S_{1/2}$ ground states of a trapped $^{171} \mathrm{Yb}^{+}$ ion, with $|0\rangle \equiv |^{2}S_{1/2},F=1,m_{F}=0\rangle $ and $|1\rangle \equiv |^{2}S_{1/2},F=1,m_{F}=1\rangle $. $|a\rangle \equiv |^{2}S_{1/2},F=0,m_{F}=0\rangle$ is an auxiliary state, as shown in Fig.~\ref{fig:device}(a). The holonomic quantum gates are realized by applying two microwave fields with time-dependent amplitude and phase, which are resonantly coupled to the transitions $|0\rangle\leftrightarrow|a\rangle$ and $|1\rangle\leftrightarrow|a\rangle$, respectively. In the interaction picture with respect to the free Hamiltonian of the ion, this interaction Hamiltonian can be written as
\begin{align}\label{initialH}
H_1(t) &= \frac{\Omega _0(t)}{2}e^{-i\phi_0(t)}|0\rangle \langle a|+\frac{\Omega _1(t)}{2}e^{-i\phi_1(t)}|1\rangle \langle a|+\mathrm{H.c.}  \notag\\
&=\frac{\Omega (t)}{2}e^{-i\phi_0(t)} |b\rangle\langle a|+\mathrm{H.c.}
\end{align}
Here, $\Omega_j$ and $\phi_j$ ($j=0,1$) are related to the time-dependent amplitudes and phases, respectively,  of the two microwave fields; $\Omega(t) = \sqrt{\Omega_0^2(t)+\Omega_1^2(t)}$ and $|b\rangle=\sin(\theta/2)|0\rangle -\cos(\theta/2) e^{i\phi}|1\rangle$, where $\tan(\theta/2) = \Omega_0(t)/\Omega_1(t)$, with $\theta$ being time-independent, and $\phi =\phi_0(t)-\phi_1(t)+\pi$ is a constant angle. From the Hamiltonian $H_1(t)$, the quantum dynamics reduces to a resonant coupling between the bright state $|b\rangle$ and  the auxiliary state $|a\rangle$, which leaves the dark state $|d\rangle=-\cos(\theta/2)e^{-i\phi}|0\rangle-\sin(\theta/2)|1\rangle$  decoupled from the $\{|b\rangle, |a\rangle\}$ subspace. That is, under the Hamiltonian $H_1(t)$ with constant $\theta$ and $\phi$, the dark state $|d\rangle$ will remain  unchanged. Thus, all of the  dynamical processes  occur   within the $\{|b\rangle,|a\rangle\}$ subspace, in which  the time-dependent Schr\"{o}dinger equation $i(\partial/\partial t)|\psi(t)\rangle=H_1(t)|\psi(t)\rangle$ is satisfied \cite{Daems13}. In this subspace, the evolution  state $|\psi(t)\rangle$ that induces the holonomy can generally be parameterized by two time-dependent angles $\alpha(t)$ and $\beta(t)$ and a global time-dependent phase $f(t)$ as follows:
\begin{equation}\label{state}
|\psi(t)\rangle = e^{-if(t)/2}\begin{pmatrix}
\cos[\alpha(t)/2] e^{i\beta(t)/2}\\[6pt]
\sin[\alpha(t)/2] e^{-i\beta(t)/2} 
\end{pmatrix}.
\end{equation}
According to the time-dependent Schr\"{o}dinger equation, the relationships between the parameters  in the evolution state $|\psi(t)\rangle$ and the Hamiltonian $H_1(t)$ can be solved as
\begin{subequations}\label{relation}
\begin{align}
\dot{f}(t)&=\dot {\beta}(t)/\cos\alpha(t),\\
\dot{\alpha}(t)&=\Omega(t) \sin{[\beta(t)+\phi_0(t)]}, \\
\dot{\beta} (t)&= \Omega(t) \cot{\alpha(t)}\cos{[\beta(t)+\phi_0(t)]},
\end{align}
\end{subequations}
where the dot indicates the time derivative. In particular, under the conditions in Eqs.~(\ref{relation}), we can choose an appropriate set of variables $\alpha(t)$, $f(t)$, and $\beta(t)$ to inversely engineer the Hamiltonian $H_1(t)$ and determine a target evolution path. Therefore, we can design the path to induce a pure non-Abelian geometric phase on the bright state $|b\rangle$ after a cyclic evolution \cite{Liu19,Li20}, from which arbitrary nonadiabatic holonomic single-qubit quantum gates can be constructed in the $\{|0\rangle, |1\rangle\}$ subspace.

Specifically, during a cyclic evolution with time $T$, we set $\alpha(t) =\pi \sin^{2}(\pi t /T)$ and $f(t) =\eta[2 \alpha - \sin (2 \alpha)]$, with $\eta$ being a constant, which chooses an evolution path for the purpose of  optimization; see Appendix~\ref{appA} for details. Meanwhile, the evolution path should be divided into two equal time intervals $[0,T/2]$ and $[T/2,T]$. During the first interval $t\in[0,T/2]$, the initial value of $\beta(t)$ is set to $\beta_1(0)=0$, and $\beta_1(T/2)=\int_0^{T/2}\dot{f}(t)\cos\alpha(t)\,dt=0$. The corresponding  evolution operator is $U_1(T/2,0) = |d\rangle \langle d| + e^{i\gamma_1}|a\rangle \langle b|$, where $\gamma_1 = -f(T/2)/2$. During the second interval $t\in[T/2,T]$, we set $\beta_2(T/2)=\gamma$, and $\beta_2(T)=\int_{T/2}^{T}\dot{f}(t)\cos\alpha(t)\,dt=\gamma$, with $\gamma$ being an arbitrary constant angle. Then, the evolution operator is $U_2(T,T/2) = |d\rangle \langle d| + e^{i\gamma_2}|b\rangle \langle a|$, where $\gamma_2 = f(T/2)/2 + \gamma$. For geometric visualization of the cyclic evolution,  as shown in Fig.~\ref{fig:device}(b), the two evolution paths have rotational symmetry on the Bloch sphere, and the cyclic geometric phase is exactly the rotation angle $\gamma$ corresponding to half  the solid angle of the rotation area. Therefore, even if   dynamical phase has accumulated during the evolution process, it will be concealed  at the end of the cyclic evolution. Moreover, the dark state is always decoupled. Consequently, the holonomic matrix is given by $U(T,0) = |d\rangle \langle d| + e^{i\gamma}|b\rangle \langle b|$ in the $\{|d\rangle,|b\rangle\}$ subspace, as a consequence of which  arbitrary holonomic single-qubit gates in the qubit basis $\{|0\rangle,|1\rangle\}$ are represented as
\begin{equation}\label{Ua}
U(\theta,\phi,\gamma)=e^{i\gamma/2}e^{-i(\gamma/2)\mathbf{n}\cdot \bm{\upsigma}},
\end{equation}
where $\mathbf{n}=(\sin\theta\cos\phi,\sin\theta\sin\phi,\cos\theta)$ and the components of $\bm{\upsigma}$ are the Pauli matrices. This rotation matrix $U$ describes a rotation  around the axis $\mathbf{n}$ by an angle $\gamma$, up to a global phase factor $\exp(i\gamma/2)$.

\section{Experimental realizations}
Our experiment is performed on a trapped $^{171}\mathrm{Yb}^{+}$ ion, with the simplified circuits being shown schematically  in Fig.~\ref{fig:device}(c). The two energy-level differences between $|0\rangle, |a\rangle$, and $|0\rangle, |1\rangle$ in our qutrit are  measured to be $\omega_{0a}=12.6428$~GHz and $\omega_{01}=12.5$~MHz, with the corresponding magnetic field being about 8.93~G. This magnetic field is produced by 30 permanent magnets fixed in a circular aluminum holder to avoid  magnetic disturbances \cite{ruster2016long}. The microwave driving on the qutrit is generated from a 12.4428~GHz signal generator (Agilent E8257D), and this is mixed with a 200~MHz microwave signal generated from an arbitrary waveform generator (AWG). After a high-pass filter (HPF), this signal is amplified to about 10~W and then sent to the ion via a microwave horn \cite{cui2016experimental}. Our trap device is shielded with a 1.5-mm-thick single layer of mu-metal \cite{farolfi2019design}, making  the coherence times, measured by Ramsey experiments, about 20~ms and 200~ms for the $|1\rangle \leftrightarrow |a\rangle$  and $|0\rangle \leftrightarrow |a\rangle $ transitions, respectively. Through optimized time-dependent amplitude and phase modulation, we can couple the transitions $|0\rangle \leftrightarrow |a\rangle$ and $|1\rangle \leftrightarrow |a \rangle$ simultaneously, to achieve  a target geometric gate.

In each cycle, the experiment adopts the following procedure: after 1~ms Doppler cooling, the state of the ion is initialized to $|a\rangle $ by 20~$\upmu$s optical pumping \cite{olmschenk2007manipulation}. Then, a resonant microwave drives the transition  $|a\rangle\leftrightarrow|0\rangle$, which prepares  the ion in the $|0\rangle$ state with $99.5\%$ fidelity. The holonomic gates are obtained through two microwave fields with frequencies $\omega_{0a}$ and $\omega_{1a}$ resonantly coupled to the $|0\rangle \leftrightarrow |a\rangle$ and $|1\rangle \leftrightarrow |a \rangle$ transitions, the amplitudes and phases of  which are modulated according to the conditions in Eqs.~(\ref{relation}). Finally, a microwave field with frequency $\omega_{0a}$ ($\omega_{1a}$) is resonantly coupled to the transition $|0\rangle \leftrightarrow |a\rangle$ ($|1\rangle \leftrightarrow |a \rangle$), which transfers the final $|0\rangle $ ($|1\rangle$) state to the $|a\rangle $ state for the purpose of  state-dependent fluorescence detection, through an objective with numerical aperture (NA) = 0.4.

\begin{figure}
\includegraphics[width=8cm]{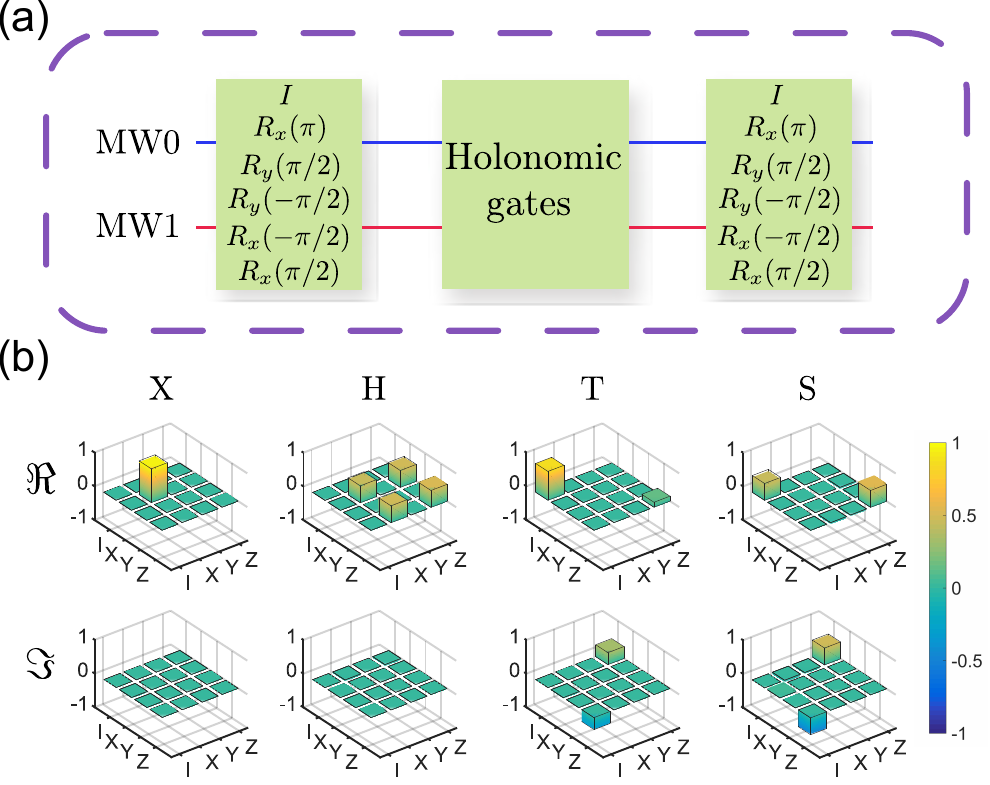}
\caption{QPT results for holonomic single-qubit quantum gates. (a) Experimental sequences for the QPT. A group of complete bases are used to prepare the initial state, and each final state is reconstructed by performing quantum state tomography. An arbitrary holonomic single-qubit gate to be characterized is implemented in the middle of these sequences. (b) Bar charts of the real and imaginary parts of the process matrix for the $X$, $H$, $T$, and $S$ gates. The solid black outlines  are plotted from the theoretical estimation.}
\label{fig:QPT}
\end{figure}

Here, we first describe the experiment with the global time-dependent phase being taken as $f(t) =\eta[2 \alpha - \sin (2 \alpha)]$; without loss of generality, we choose  $\eta = 1/5$. The corresponding gate duration is 224~$\upmu$s under the maximum driving strength of $(2\pi)$ 10 kHz. We characterize the single-qubit holonomic gate through standard quantum process tomography (QPT),  and the corresponding experimental sequences are shown in Fig.~\ref{fig:QPT}(a). In the QPT process, we first prepare a set of states $\{ |0\rangle, |1\rangle, (|0\rangle + |1\rangle )/\sqrt{2}, (|0\rangle - |1\rangle )/ \sqrt{2}, (|0\rangle + i|1\rangle )/ \sqrt{2},  (|0\rangle - i|1\rangle )/ \sqrt{2} \}$ by applying the set of respective operations $\{ I, R_{x}(\pi), R_{y}(\pi/2), R_{y}(-\pi/2), R_{x}(-\pi/2), R_{x}(\pi/2) \}$ to the state $|0\rangle$. 
The holonomic gate is then implemented immediately. Finally, we measure the output states through quantum state tomography to reconstruct the states.
The process matrices are estimated from all the results through a maximum likelihood estimation method \cite{jevzek2003quantum}. We use the process fidelity  $F_\mathrm{att}=|\mathrm{Tr}(\chi_\mathrm{exp} \chi_\mathrm{th}^{\dagger})|$ to evaluate the QPT results, where $\chi_\mathrm{exp}$ and $\chi_\mathrm{th}$ are the experimental and theoretical process matrices, respectively. The experimental results for four example gates $U(\pi/2,0,\pi)$, $U(\pi/4,0,\pi)$, $U(0,0,\pi/4)$, and $U(0,0,\pi/2)$, which are respectively the $X$, $H$, $T$, and $S$ gates, are shown in Fig.~\ref{fig:QPT}(b), and the corresponding gate fidelities are found to be $F_{X}=97.21\% \pm 0.03\%$, $F_{H}=97.65\% \pm 0.06\%$, $F_{T}=97.85\% \pm 0.05\%$, and $F_{S}=97.43\% \pm 0.03\%$.

\begin{figure}
\includegraphics{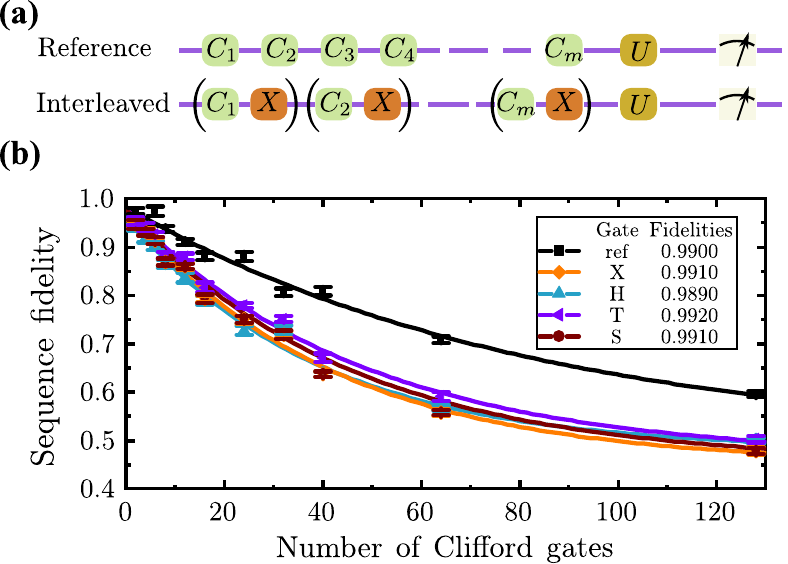}
\caption{ RB  results for holonomic single-qubit quantum gates. (a) Sequences of reference and interleaved RB experiments. (b) Sequence fidelity with the number of Clifford gates.  The fidelity for each sequence length is measured for 20 different random sequences, with the standard deviation from the mean plotted as the error bars. All curves are fitted through $F=Ap^{m}+B$. The average gate fidelity is calculated using $F_\mathrm{ave}=1-(1-p_\mathrm{ref})/2$, and a specific gate fidelity is calculated using $F_\mathrm{gate}=1-(1-p_\mathrm{gate}/p_\mathrm{ref})/2$, where $p_\mathrm{ref}$ and $p_\mathrm{gate}$ are the $p$ parameters for the reference and interleaved RB processes, respectively.}
\label{fig:RB} 	
\end{figure}

To characterize the performance of the implemented holonomic single-qubit quantum gates, we also use the random benchmarking (RB) method, which is not dependent on perfect state preparation and measurement. A reference RB experiment and an interleaved RB experiment are performed to investigate the fidelity of the implemented  gates, the experimental sequences of which are shown in Fig.~\ref{fig:RB}(a). The results for four holonomic single-qubit gates are shown in Fig.~\ref{fig:RB}(b). The reference RB experiment gives the average fidelity $F_\mathrm{ave}=99\%$ of single-qubit gates in the Clifford group. The interleaved RB experiment gives the fidelity of specific holonomic gates, which are found to be $F_{X}=99.10\%$, $F_{H}=98.90\%$, $F_{T}=99.20\%$, and $F_{S}=99.10\%$, respectively. The remaining infidelity arises mainly from decoherence of the  $|1\rangle$ state due to  magnetic field disturbance.

\section{robustness test}
\begin{figure}	
\includegraphics{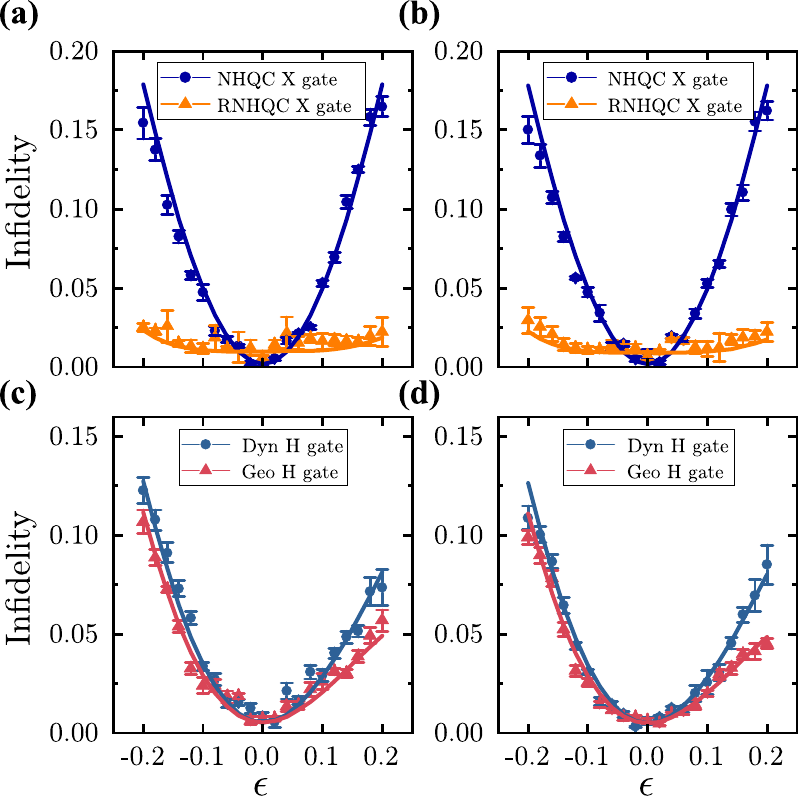}
\caption{Noise-resilient features of different  types of quantum gates. In (a) and (b), the  $X$ and $H$ gate infidelities, respectively, are plotted as functions of the Rabi frequency error for  RNHQC with $\eta = 1$ and conventional NHQC  with $\eta = 0$.  These results indicate the superiority of RNHQC  against Rabi frequency error, especially when the error is large. In (c) and (d), the infidelities of both holonomic and dynamical gates are plotted as functions of the Rabi frequency error for $X$ and $H$ gates, respectively. These results indicate the superior robustness  of holonomic gates compared with dynamical ones. The error bars indicate the standard deviation, and each data point is averaged over 2000 realizations.}
	\label{fig:error}
\end{figure}

We now proceed to demonstrate the  robustness of the gates against control amplitude errors, which are among the main sources of gate errors in large-scale quantum systems. We compare RNHQC gates ($\eta=1$) with  conventional NHQC gates ($\eta=0$) (see Appendix~\ref{appA} for details), under the same maximum  driving strength [$(2\pi)$10~kHz in our experiment], which is limited by the power of our amplifier. The gate durations of the RNHQC and NHQC gates are 815.6~$\upmu$s and 157~$\upmu$s, respectively. Figures~\ref{fig:error}(a) and \ref{fig:error}(b) show the results of our experimental characterization of the performances of the $X$ and $H$ gates, respectively, as  functions of the Rabi frequency error $\epsilon$, obtained using a single-qubit RB method for both the RNHQC and NHQC cases. All of the experimental results  agree very well with the numerical simulations. The comparisons clearly illustrate the distinct advantages of the realized RNHQC gates with regard to the effect of amplitude errors, especially in  the case of large errors, for both of the demonstrated gates.

We also demonstrate the superiority of quantum gates in RNHQC  over dynamical gates in terms of robustness against amplitude errors, under the same  pulse shape. In our experiment, we compare the holonomic gates $(\eta)$ with the dynamical gates $(\eta_D)$ under the same driving pulse, with the maximum driving strength being $(2\pi)$10~kHz and $\eta=\eta_D$ (see Appendix~\ref{appB} for details). The corresponding durations of the geometric and dynamical gates are both 429.8~$\upmu$s. Figures~\ref{fig:error}(c) and \ref{fig:error}(d) show the results of our experimental characterization of the performances of the $X$ and $H$ gates, respectively, as functions of the Rabi frequency error $\epsilon$, obtained using the single-qubit RB method for both the RNHQC and dynamical cases.  These results again show the superiority of holonomic gates over  dynamical ones.

\section{nontrivial two-qubit gates}
To achieve  universal quantum computation, two-qubit entangling gates are also necessary. Here, we propose a feasible scheme to implement a robust nonadiabatic holonomic controlled-phase gate between the internal atomic states $\{|0\rangle,|1\rangle\}$ and the motional state of the ion. Only the $\{|0\rangle,|1\rangle\}$ subspace of the motional state is considered, while its Hilbert space is infinite. To avoid  leakage out of the logical subspace, a resonant blue sideband drive couples the $|a0\rangle \leftrightarrow |11\rangle$ transition (see Appendix~\ref{appC} for details), with the effective Hamiltonian in the interaction picture being
\begin{equation}\label{H2}
H_{2}(t)=\frac{\tilde{\Omega}(t)}{2}e^{-i\tilde{\phi}(t)}|11\rangle \langle a0|+\frac{\tilde{\Omega}(t)}{2}e^{i\tilde{\phi}(t)}|a0\rangle \langle 11|,
\end{equation}
where $\tilde{\Omega}(t)$ and  $\tilde{\phi}(t)$  are the effective coupling strength and phase of the parametric drive, respectively. This effective Hamiltonian has the same form as  the Hamiltonian for the single-qubit case  in Eq.~\eqref{initialH}, with $\theta = 0$, and now the corresponding bright state is $|b\rangle = |11\rangle$. Similarly to the single-qubit case, a geometric gate ${\mathrm{diag}} (e^{i\gamma},e^{-i\gamma})$ in the subspace $\{|a0\rangle,|11\rangle\}$ can be realized through modulating the effective coupling strength $\tilde{\Omega}(t)$ and its phase $\tilde{\phi}(t)$. For the two-qubit computational subspace of $\{|00\rangle,|01\rangle,|10\rangle,|11\rangle\}$, the resulting geometric operation reduces to a controlled-phase gate with a conditional phase $\gamma$, i.e.,
\begin{equation}\label{U2}
U(\gamma)=\mathrm{diag}(1,1,1,e^{i\gamma}).
\end{equation}
A universal set of  quantum gates can then be realized in this trapped-ion system.

\section{conclusion}
In conclusion, we have  experimentally demonstrated arbitrary robust nonadiabatic holonomic single-qubit gates with resonant drives. The superiority against control amplitude error of these gates has been verified by comparing them with  gates in conventional NHQC, as well as with dynamical gates. This distinct advantage of holonomic gates illustrates their promise as candidates for future robust quantum computation. Finally, aiming at a universal RNHQC, we have also proposed a scheme for a nontrivial two-qubit control phase gate, which can be realized with the spin qubit and phonon qubit of an ion.  Thus, our work confirms the feasibility of RNHQC using trapped ions.

\acknowledgments
This work was supported by the National Key Research and Development
Program of China (Nos. 2017YFA0304100, 2016YFA0302700, and 2016YFA0301803), the National
Natural Science Foundation of China (Nos. 11874343, 11774335, 11734015, 11821404, and 11874156), the
Anhui Initiative in Quantum Information Technologies (AHY020100 and AHY070000), the
Key Research Program of Frontier Sciences,	CAS (No. QYZDY-SSW-SLH003), the Science Foundation of the CAS (No. ZDRW-XH-2019-1), and the Fundamental Research Funds for the Central Universities (Nos. WK2470000026, WK2470000027, and WK2470000028).

\begin{figure}
	\includegraphics[width=8cm]{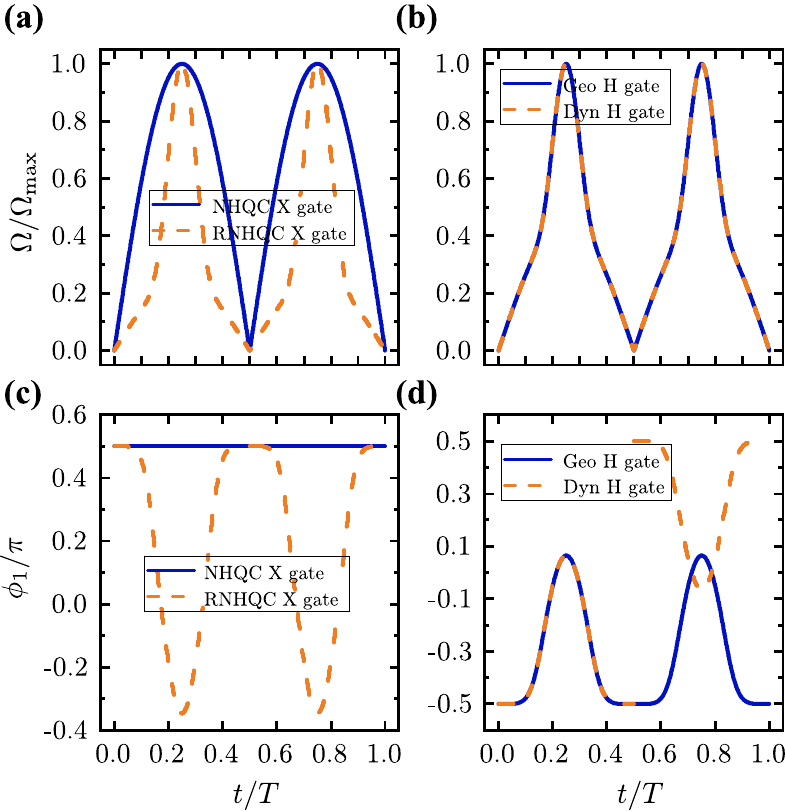}
	\caption{Amplitudes and phases of the microwave field used in our experiments. (a) and (c) show the normalized amplitudes and phases, respectively, of  conventional NHQC and RNHQC X gates. (b) and (d) show the normalized amplitudes and phases, respectively, of RNHQC  and dynamical H gates.}\label{fig:wave} 	
\end{figure}

\appendix

\section{Optimal control technique}\label{appA}

For the Hamiltonian $H_1(t)$ in Eq.~(\ref{initialH}), there are two adjustable time-dependent parameters, which enables us to incorporate optimal control technique \cite{Chen12,Daems13} to further enhance the robustness of the implemented holonomic quantum gates against amplitude errors.
Here, we consider the static amplitude error situation, i.e., $\Omega(t)\rightarrow(1+\epsilon)\Omega(t)$, and the Hamiltonian with this error can then be written as
\begin{equation}\label{HE}
H_{\epsilon}(t) =(1+\epsilon)\frac{\Omega(t)}{2} e^{-i\phi_0(t)} |b\rangle\langle a| + \mathrm{H.c.}
\end{equation}
Under this type of error, an optimal control technique can be adopted during the two intervals for the single-loop evolution to induce the holonomic quantum gates. Specifically, the influence of the  error can be evaluated at the end of the first interval $T/2$, and the probability amplitude  is then given by $P_\epsilon=| \langle \psi(\tau/2)|\psi_\epsilon(\tau/2)\rangle|^2
= 1 + \tilde{O}_1+\tilde{O}_2+\cdots$,
where $ |\psi_\epsilon(\tau/2)\rangle $ is the output state under the static amplitude error, and $\tilde{O}_m$ is the $m$th-order perturbation term. For realistic experimental realization, up to the second order, this probability amplitude reads
\begin{equation}\label{Qs}
P_\epsilon \simeq 1-\epsilon^2
\left|\int^{T/2}_0 e^{-if}\dot{\alpha}\sin^2{\alpha}\, dt \right|^2.
\end{equation}
To achieve $| \langle \psi(\tau/2)|\psi_\epsilon(\tau/2)\rangle|^2
\simeq 1$, we set $\alpha(t) =\pi \sin ^{2}(\pi t /T)$, $f(t) =\eta[2 \alpha - \sin (2 \alpha)]$, $\beta_1(0)=0$, and $\beta_1(T/2)=\gamma$, which gives
\begin{equation}
P_\epsilon \simeq 1-\epsilon^2\sin^2\eta\pi/(2\eta)^2;
\end{equation}
i.e., for nonzero  integer $\eta$, $P_{\epsilon} \simeq 1$. When $\eta\rightarrow0$, the current implementation reduces to the previous NHQC case.  For experimental realization, comparing with the previous NHQC case with $\eta=0$, we select $\eta=1$ to demonstrate the robustness of nonadiabatic holonomic quantum gates against amplitude errors within the range of $-0.2\leq\epsilon\leq0.2$. To provide a fair comparison, the same maximum value of $\Omega(t)$ is set  in both cases, namely, $\Omega_\mathrm{max}= (2\pi) 10$~kHz. The maximum value of the optimized pulse is bounded by $\Omega_\mathrm{max}$, and thus the improvement in  gate performance can only be attributed to the optimal control. Specifically, the time-dependent amplitudes and phases of both microwave fields in our experiment with $\eta=0$ and $\eta=1$ for the X gate $U(\pi/2,0,\pi)$ are given in Figs.~\ref{fig:wave}(a) and \ref{fig:wave}(c).

\section{Dynamical gate}\label{appB}
Here, we present the construction of arbitrary quantum gates without canceling  dynamical phase in a single-loop evolution. Specifically, a cyclic evolution with time $T$ is divided into two equal time intervals $[0,T/2]$ and $[T/2,T]$. During the first interval $t\in[0,T/2]$, we set $\alpha(t) =\pi \sin^{2}(\pi t /T)$ and $f(t) =\eta_D[2 \alpha - \sin (2 \alpha)]$. The initial value of $\beta(t)$ is set  to $\beta_1(0)=0$, and $\beta_1(T/2)=\int_0^{T/2}\dot{f}(t)\cos\alpha(t)\,dt=0$. The corresponding  evolution operator is $U_1(T/2,0) = |d\rangle \langle d| + e^{i\gamma_1}|a\rangle \langle b|$, where $\gamma_1 = -f_1(T/2)/2$. During the second interval $t\in[T/2,T]$, we set $\alpha(t) =\pi \sin ^{2}(\pi t /T)$, and $f(t) =-\eta_D[2 \alpha - \sin (2 \alpha)]$, $\beta_2(T/2)=0$, and $\beta_2(T)=\int_{T/2}^{T}\dot{f}(t)\cos\alpha(t)\,dt=0$. The  evolution operator is then $U_2(T,T/2) = |d\rangle \langle d| + e^{i\gamma_2}|b\rangle \langle a|$, where $\gamma_2 = -f_1(T/2)/2$. Overall, after this cyclic evolution, the evolution matrix is given by $U(T,0) = |d\rangle \langle d| + e^{i\gamma_D}|b\rangle \langle b|$ in the $\{|d\rangle,|b\rangle\}$ subspace with $\gamma_D =-f_1(T/2)=-2\eta_D\pi$, and arbitrary dynamical single-qubit gates are then represented in the qubit basis $\{|0\rangle,|1\rangle\}$ as
\begin{equation}\label{Ua1}
U(\theta,\phi,\gamma_D)=e^{i\gamma_D/2}e^{-i(\gamma_D/2)\mathbf{n}\cdot \bm{\upsigma}}.
\end{equation}

\begin{figure}
\includegraphics{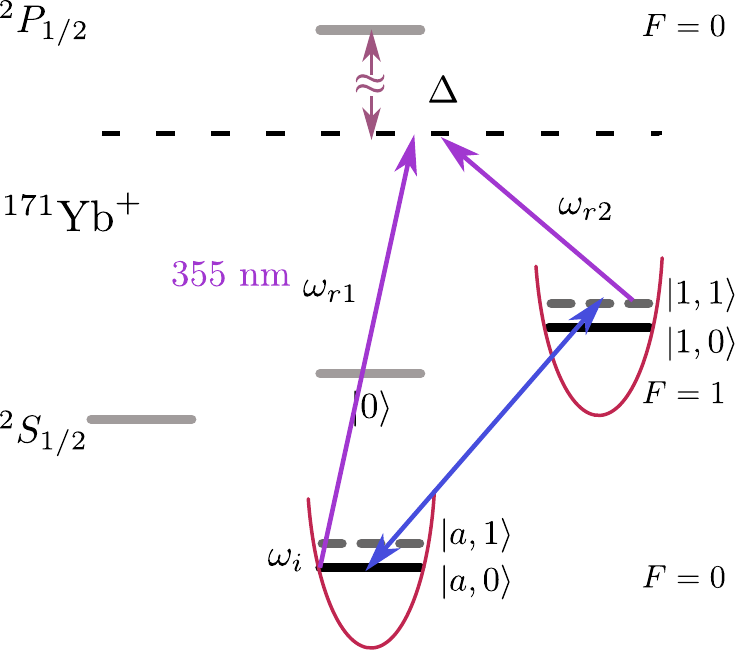}\caption{Energy level diagram of a trapped $^{171}\mathrm{Yb}^{+}$
ion. Two 355~nm pulsed laser beams perpendicular to each other are
used to excite the motional mode. The difference in frequency between
these two lasers is $\omega_{r1}-\omega_{r2}=\omega_{0a}+\omega_{x}$,
which corresponds to a blue sideband operation. By optimizing the waveform
of this blue sideband, a nontrivial two-qubit phase gate can be realized
as mentioned in the main text.}\label{fig6}
\end{figure}

We can then experimentally confirm the superiority of the geometric phase gate over the dynamical one in terms of robustness against control amplitude errors $-0.2\leq\epsilon\leq0.2$ under the same driving amplitude. Because the dynamical evolution depends on the parameter $\eta_D$, we can construct  geometric gates for the same evolution matrix as in the main text under the condition $\eta_D = \eta$ for the same driving amplitude and $\gamma_D =-2\eta_D\pi=\gamma$ and with the same parameters $(\theta, \phi)$. The time-dependent amplitudes and phases of our experimental microwave fields with $\eta_D = \eta =-1/2$ for the H gate $U(\pi/4,0,\pi)$ are given in Figs.~\ref{fig:wave}(b) and \ref{fig:wave}(d).

\section{Coupling of spin and phonons}\label{appC}

The motion of a charged particle trapped in a Paul trap can be described
by a set of basis states $|n\rangle $ in which $n=1,2,\dots,\infty$
\citep{leibfried2003quantum}. These so-called phonon states are the
dynamical counterpart of the harmonic-oscillator number (Fock) state.
A nontrivial two-qubit control phase gate can be implemented on
the spin and motional state of an ion. As shown in Fig.~\ref{fig6}, the interaction
between the spin with energy splitting $\omega_{0a}$ and the phonons
with frequency $\omega_{i}$ can be induced by a pair of appropriately selected
stimulated Raman laser beams with beat-note frequency $\omega_{r1}-\omega_{r2}=\omega_{0a}\pm\omega_{i}$,
in which $+(-)$ denotes the blue (red) sideband \cite{leibfried2003experimental}.
The blue sideband interaction can be described by the effective two-level anti-Jaynes--Cummings Hamiltonian
\begin{equation}
H_{\text{anti-JC}}=i\hbar\Omega_{r}\eta_{\textrm{LD}}(\sigma_{+}a^{\dagger}e^{i\phi}+\mathrm{H.c.})
\end{equation}
under the rotational wave approximation (RWA), where $\Omega_{r}=\sqrt{\Omega_{1}\Omega_{2}}$
is the total Rabi frequency, $\eta_{\textrm{LD}}$ is the Lamb--Dicke coefficient, and
$a^{\dagger}~(a)$ is the creation (annihilation) operator. To obtain
tunable coupling, we can add an arbitrary waveform generator (AWG)
to modulate the phase $\phi$ and amplitude $\Omega_{r}$
of one beam through an acousto-optical modulator (AOM). Just like the microwave operations  mentioned
in the main text, with optimized phase and amplitude,  two-qubit
control phase gates could be realized through the blue sideband transition
 $|a,0\rangle \leftrightarrow|1,1\rangle $. 

On our platform,  $\eta_{\textrm{LD}}=0.1$ and $\omega_{x}=2.4$~MHz. With 20~mW average power in each laser beam, the effective Rabi
frequency $\eta_{\textrm{LD}}\Omega_{r}$ of the blue sideband is about 100~$\upmu$s.
The corresponding time of the two-qubit gate is about 224~$\upmu$s with
the optimized parameter $\eta=1/5$. Although Raman operation will introduce more errors than microwave operation, such as disturbances of the intensity and phase of the laser, the stimulated Raman process can be manipulated very rapidly with sufficient laser power. Besides, the single-qubit gates experimentally demonstrated in this paper could also be
realized through the stimulated Raman process with  frequency $\omega_{r1}-\omega_{r2}=\omega_{0a}$.
With arbitrary single-qubit gates and a two-qubit control phase gate,
 universal quantum computation could be realized on a trapped-ion platform.


\begin{thebibliography}{99}
\bibitem{berry84} M. V. Berry, 
Quantal phase factors accompanying adiabatic changes, 
Proc. R. Soc. Lond. A {\bf 392}, 45 (1984).

\bibitem{zee84} F. Wilczek and A. 
Zee, Appearance of gauge structure in simple dynamical systems,
Phys. Rev. Lett.  {\bf 52}, 2111 (1984).



\bibitem{ps04} P. Solinas, P. Zanardi, and N. Zangh\`{\i},
Robustness of non-Abelian holonomic quantum gates against parametric noise,
Phys. Rev. A {\bf 70}, 042316 (2004).

\bibitem{zhu05} S.-L. Zhu and P. Zanardi,
Geometric quantum gates that are robust against stochastic control errors,
Phys. Rev. A {\bf 72}, 020301(R) (2005).

\bibitem{ps12} P. Solinas, M. Sassetti, T. Truini, and N. Zangh\`{\i},
On the stability of quantum holonomic gates,
New J. Phys. {\bf 14}, 093006 (2012).

\bibitem{mj12} M. Johansson, E. Sj\"{o}qvist, L. M. Andersson, M. Ericsson, B. Hessmo, K. Singh, and D. M. Tong, 
Robustness of nonadiabatic holonomic gates, 
Phys. Rev. A {\bf 86}, 062322 (2012).

\bibitem{AA87} Y. Aharonov and J. Anandan,
Phase change during a cyclic quantum evolution,
Phys. Rev. Lett. {\bf 58}, 1593 (1987).



\bibitem{xbwang01} WangXiang-Bin and M. Keiji
Nonadiabatic conditional geometric phase shift with NMR,
Phys. Rev. Lett. {\bf 87}, 097901 (2001).

\bibitem{zhu02} S.-L. Zhu and Z. D. Wang,
Implementation of universal quantum gates based on nonadiabatic geometric phases,
Phys. Rev. Lett. {\bf 89}, 097902 (2002).

\bibitem{gqc08} E. Sj\"{o}qvist,
A new phase in quantum computation,
Physics {\bf1}, 35 (2008).



\bibitem{zanardi99} P. Zanardi and M. Rasetti,
Holonomic quantum computation,
Phys. Lett. A {\bf 264}, 94 (1999).

\bibitem{Duan01} L. M. Duan, J. I. Cirac, and P. Zoller, 
Geometric manipulation of trapped ions for quantum computation
Science \textbf{292}, 1695 (2001).

\bibitem{toyoda2013realization} K. Toyoda, K. Uchida, A. Noguchi, S. Haze, and S. Urabe,
Realization of holonomic single-qubit operations,
Phys. Rev. A  {\bf 87}, 052307 (2013).

\bibitem{leroux2014} F. Leroux, K. Pandey, R. Rehbi, F. Chivy, C. Miniatura, B. Gr\'{e}maud, and D. Wilkowski,
Non-Abelian adiabatic geometric transformations in a cold Strontium gas,
Nat. Commun. \textbf{9}, 3580 (2018).

\bibitem{Sjoqvist2012}
E.~Sj\"{o}qvist, D.~M. Tong, L.~{Mauritz Andersson}, B.~Hessmo, M.~Johansson, and K.~Singh,
Non-adiabatic holonomic quantum computation,
New J. Phys. \textbf{14}, 103035 (2012).

\bibitem{Xu2012} G.~F. Xu, J.~Zhang, D.~M. Tong, E.~Sj\"{o}qvist, and L.~C. Kwek,
Nonadiabatic holonomic quantum computation in decoherence-free subspaces,
Phys. Rev. Lett. \textbf{109}, 170501 (2012).









\bibitem{Xu2015} G.~F. Xu, C.~L. Liu, P.~Z. Zhao, and D.~M. Tong,
Nonadiabatic holonomic gates realized by a single-shot implementation,
Phys. Rev. A   \textbf{92}, 052302 (2015).


\bibitem{es2016} E. Sj\"{o}qvist,
Nonadiabatic holonomic single-qubit gates in off-resonant $\Lambda$ systems,
Phys. Lett. A {\bf 380}, 65 (2016).



\bibitem{Herterich2016} E.~Herterich and E.~Sj\"{o}qvist,
Single-loop multiple-pulse nonadiabatic holonomic quantum gates,
Phys. Rev. A \textbf{94}, 052310   (2016).

\bibitem{xu2017} G. F. Xu, P. Z. Zhao, T. H. Xing, E. Sj\"{o}qvist, and D. M. Tong,
Composite nonadiabatic holonomic quantum computation,
Phys. Rev. A {\bf 95}, 032311 (2017).

\bibitem{Xue2017} Z.-Y. Xue, F.-L. Gu, Z.-P. Hong, Z.-H. Yang, D.-W. Zhang, Y. Hu, and J. Q. You,
Nonadiabatic holonomic quantum computation with dressed-state qubits,
Phys. Rev. Appl. {\bf 7}, 054022 (2017).


\bibitem{xu20172} P. Z. Zhao, G. F. Xu, Q. M. Ding, Erik Sj\"{o}qvist, and D. M. Tong,
Single-shot realization of nonadiabatic holonomic quantum gates in decoherence-free subspaces,
Phys. Rev. A {\bf 95}, 062310 (2017).


%

\bibitem{Hong18} Z.-P. Hong, B.-J. Liu, J.-Q. Cai, X.-D. Zhang, Y. Hu, Z.-D. Wang, and Z.-Y. Xue,
Implementing universal nonadiabatic holonomic quantum gates with transmons,
Phys. Rev. A {\bf 97}, 022332 (2018).

\bibitem{xu2018} G. F. Xu, D. M. Tong, and E. Sj\"{o}qvist,
Path-shortening realizations of nonadiabatic holonomic gates,
Phys. Rev. A {\bf 98}, 052315 (2018).


\bibitem{nr2019} N. Ramberg and E. Sj\"{o}qvist, Environment-Assisted Holonomic Quantum Maps,
Phys. Rev. Lett. {\bf 122}, 140501 (2019).






\bibitem{Abdumalikov2013} A.~A. Abdumalikov, J.~M. Fink, K.~Juliusson, M.~Pechal, S.~Berger, A.~Wallraff, and S.~Filipp, Experimental realization of non-Abelian non-adiabatic geometric gates, 
Nature (London) \textbf{496}, 482 (2013).

\bibitem{Xu18}
Y. Xu, W. Cai, Y. Ma, X. Mu, L. Hu, T. Chen, H. Wang, Y.-P. Song, Z.-Y. Xue, Z.-Q. Yin, and L. Sun,
Single-loop realization of arbitrary nonadiabatic holonomic single-qubit quantum gates in a superconducting circuit,
Phys. Rev. Lett. {\bf 121}, 110501 (2018).
%
\bibitem{Danilin18}
S. Danilin, A. Veps\"{a}l\"{a}inen, and G. S. Paraoanu,
Experimental state control by fast non-Abelian holonomic gates with a superconducting qutrit, 
Phys. Scripta {\bf 93}, 055101 (2018).

\bibitem{Feng2013} G.~Feng, G.~Xu, and G.~Long,
Experimental realization of nonadiabatic holonomic quantum computation,
Phys. Rev. Lett. \textbf{110}, 190501   (2013).

\bibitem{li2017}  H. Li, L. Yang, and G. Long,
Experimental realization of single-shot nonadiabatic holonomic gates in nuclear spins,
Sci. China: Phys. Mech. Astron. {\bf 60}, 080311(2017).

\bibitem{zhu2019} Z. Zhu, T. Chen, X. Yang, J. Bian, Z.-Y. Xue, and X. Peng,
single-loop and composite-loop nonadiabatic holonomic quantum computation in a decoherence-free subspace,
Phys. Rev. Appl. {\bf 12}, 024024 (2019).





\bibitem{Zu2014} C.~Zu, W.-B. Wang, L.~He, W.-G. Zhang, C.-Y. Dai, F.~Wang, and L.-M. Duan,
Experimental realization of universal geometric quantum gates with solid-state spins.
Nature (London) \textbf{514}, 72 (2014).   

\bibitem{Arroyo-Camejo2014} S.~Arroyo-Camejo, A.~Lazariev, S.~W. Hell, and G.~Balasubramanian,
Room temperature high-fidelity holonomic single-qubit gate on a solid-state spin,
Nat. Commun. \textbf{5}, 4870 (2014).


\bibitem{nv2017} Y. Sekiguchi, N. Niikura, R. Kuroiwa, H. Kano, and H. Kosaka,
Optical holonomic single quantum gates with a geometric spin under a zero field,
Nat. Photonics {\bf 11}, 309 (2017). 

\bibitem{nv20172}
B. B. Zhou, P. C. Jerger, V. O. Shkolnikov, F. J. Heremans, G. Burkard, and D. D. Awschalom,
Holonomic quantum control by coherent optical excitation in diamond,
Phys. Rev. Lett.  {\bf 119}, 140503 (2017).

\bibitem{ni2018}
N. Ishida, T. Nakamura, T. Tanaka, S. Mishima, H. Kano, R. Kuroiwa, Y. Sekiguchi, and H. Kosaka,
Universal holonomic single quantum gates over a geometric spin with phase-modulated polarized light,
Opt. Lett. {\bf 43}, 2380 (2018).

\bibitem{kn2018} K. Nagata, K. Kuramitani, Y. Sekiguchi, and H. Kosaka,
Universal holonomic quantum gates over geometric spin qubits with polarised microwaves,
Nat. Commun. {\bf 9}, 3227 (2018).




\bibitem{Zheng16} S.-B. Zheng, C.-P. Yang, and F. Nori, Comparison of the sensitivity to systematic errors between nonadiabatic non-Abelian geometric gates and their dynamical counterparts, 
Phys. Rev. A {\bf 93}, 032313 (2016).

\bibitem{Jing17}  J. Jing, C.-H. Lam, and L.-A. Wu,
Non-Abelian holonomic transformation in the presence of classical noise,
Phys. Rev. A {\bf 95}, 012334 (2017).


\bibitem{Liu19} B.-J. Liu, X.-K. Song, Z.-Y. Xue, X. Wang, and M.-H. Yung,
Plug-and-play approach to non-adiabatic geometric quantum computation,
Phys. Rev. Lett. {\bf 123}, 100501 (2019).

\bibitem{Li20} S. Li, T. Chen, and Z.-Y. Xue,
Fast holonomic quantum computation on superconducting circuits with optimal control
Adv. Quantum Technol. {\bf 3}, 2000001 (2019).

\bibitem{yan2019}    T. Yan, B.-J. Liu, K. Xu, C. Song, S. Liu, Z. Zhang, H. Deng, Z. Yan, H. Rong, K. Huang, M.-H. Yung, Y. Chen, and D. Yu,
Experimental realization of nonadiabatic shortcut to non-Abelian geometric gates,
Phys. Rev. Lett. {\bf 122}, 080501 (2019).


\bibitem{Daems13} D. Daems, A. Ruschhaupt, D. Sugny, and S. Gu\'{e}rin,
Robust quantum control by a single-shot shaped pulse,
Phys. Rev. Lett. {\bf 111}, 050404 (2013).


\bibitem{ruster2016long} T. Ruster, C. T. Schmiegelow, H. Kaufmann, C. Warschburger, F. Schmidt-Kaler, and U. G. Poschinger,
A long-lived Zeeman trapped-ion qubit,
Appl. Phys. B. {\bf 122}, 254 (2016).


\bibitem{farolfi2019design} A. Farolfi, D. Trypogeorgos, G. Colzi, E. Fava, G. Lamporesi, and G. Ferrari,
Design and characterization of a compact magnetic shield for ultracold atomic gas experiments,
Rev. Sci. Instrum. {\bf 90}, 115114 (2019).


\bibitem{olmschenk2007manipulation} S. Olmschenk, K. C. Younge, D. L. Moehring, D. Matsukevich, P. Maunz, and C. Monroe,
Manipulation and detection of a trapped $\mathrm{Yb}^{+}$ hyperfine qubit,
Phys. Rev. A  {\bf 76}, 052314 (2007).

\bibitem{cui2016experimental} J.-M. Cui, Y.-F. Huang, Z. Wang, D.-Y. Cao, J. Wang, W.-M. Lv, L. Luo, A. Del Campo, Y.-J. Han, and C.-F. Li,
Experimental Trapped-ion Quantum Simulation of the Kibble-Zurek dynamics in momentum space,
Sci. Rep. {\bf 6}, 33381 (2016).


\bibitem{jevzek2003quantum} M. Je\v{z}ek, J. Fiur\'{a}\v{s}ek, and Z. Hradil,
Quantum inference of states and processes,
Phys. Rev. A {\bf 68}, 012305 (2003).


\bibitem{Chen12} A. Ruschhaupt, X. Chen, D. Alonso, and J. G. Muga,
Optimally robust shortcuts to population inversion in two-level quantum systems,
New J. Phys. {\bf 14}, 093040 (2012).


\bibitem{leibfried2003quantum} D. Leibfried, R. Blatt, C. Monroe, and D. Wineland,
Quantum dynamics of single trapped ions,
Rev. Mod. Phys. {\bf 75}, 281 (2003).

\bibitem{leibfried2003experimental} D. Leibfried, B. DeMarco, V. Meyer, D. Lucas, M. Barrett, J. Britton, W. M. Itano, B. Jelenkovi\'{c}, C. Langer,  and T. Rosenband,
Experimental demonstration of a robust, high-fidelity geometric two ion-qubit phase gate,
Nature (London) {\bf 422}, 412 (2003).

\end{thebibliography}
\end{document}